\documentclass{iaus}
\usepackage{graphicx}

\title[Non-radial pulsations in Be/X binaries
] 
{Non-radial pulsations in the Be/X binaries 4U\,0115+63 and SAX\,J2103.5+4545
}

\author[J. Guti\'errez-Soto et al.]   
{J. Guti\'errez-Soto$^1,^2$, P. Reig$^3,^4$, J. Fabregat$^5$ \and L. Fox-Machado$^6$}

\affiliation{
$^1$Instituto de Astrof\'{\i}sica de Andaluc\'{\i}a (CSIC), Glorieta de la Astronom\'{\i}a s/n 18008, Granada,
Spain
email: {\tt jgs@iaa.es}\\
$^2$GEPI, Observatoire de Paris, CNRS, Universit\'e Paris Diderot, 5 place Jules Janssen, 92190 Meudon, France\\ 
$^3$IESL, Foundation for Research and Technology, 71110 Heraklion, Crete, Greece\\
$^4$University of Crete, Physics Department, PO Box 2208, 710 03 Heraklion, Crete, Greece\\
$^5$Observatorio Astron\'omico de la Universidad de
Valencia, Calle Catedr\'atico Agust\'\i n Escardino 7, 46980 Paterna, Valencia, Spain \\
$^6$ Observatorio Astron\'omico Nacional, Instituto de Astronom\'\i a – Universidad Nacional Aut\'onoma de
M\'exico, Ap. P. 877, Ensenada, BC 22860, Mexico
}

\pubyear{2010}
\volume{272}  
\pagerange{119--126}
\jname{Active OB stars: structure, evolution, mass loss and critical limits}
\editors{C. Neiner, G. Wade, G. Meynet \& G. Peters}

\newcommand{\cd}{$d^{-1}$}
\newcommand{\corot}{{\sc corot}}
\newcommand{\kepler}{{\sc kepler}}

\begin{document}

\maketitle

\begin{abstract}
The discovery of non-radial pulsations (NRP) in the Be/X binaries of the Magellanic Clouds (MC, eg. 
\cite{Fabrycky}, \cite{Coe}, \cite{Schmidtke}) provided a new approach to understand
these complex systems,
and, at the same time, favoured the synergy between two different fields: stellar pulsations and X-ray
binaries. This breakthrough was possible thanks to the MACHO and OGLE surveys. However, in our Galaxy, only
two Be/X have been reported to show NRP:  GRO\,J2058+42 (\cite{Kiziloglu}) and 
LSI+61\,235 (\cite{Sarty}). 
Our objective is to study the short-term variability of Galactic Be/X binaries, compare them to the Be/X of
the MC and to the isolated Galactic Be observed with \corot\ and \kepler. We present preliminary results of
two
Be/X stars, namely 4U\,0115+63 and SAX\,J2103.5+4545 showing multiperiodicity and periodicity respectively,
most probably produced by non-radial pulsations. 
\end{abstract}

\firstsection 
\section{Introduction}

Be/X-ray binaries consist of a neutron star orbiting a Be star. The
early-type companion is believed to have the same physical properties as an
isolated Be star. However, the structure and evolution of the equatorial
disk is affected by the presence of the compact companion
\cite{Reig07}. While there are numerous detections of NRP in
isolated Galactic Be stars, especially after the launch of the \corot\ mission
(\cite{Gutierrez2010thisproceedings}), very little work on the search of variability associated with NRP in
Galactic BeX exists. 
NRP can manifest themselves through multi-period photometric variability by
modulating the stellar surface temperature. Typical periods associated with
NRP in these stars range from few hours to 2 days.

We have set up a project to investigate the short-term optical photometric
variability of BeX in the Milky Way. The main goal of this project is to
detect NRP and compare their frequencies and amplitudes with those of
isolated Galactic Be stars and with BeX in other galaxies with different metallicity
content.

The Galactic Be/X studied in this project are 4U\,0115+63, SAX\,J2103.5+4545, AO\,0535+26 and 4U\,2206+54. 
In this paper we only show the results for the two first targets. 

%

Optical differential CCD-photometry was obtained in 3 sites in two observing campaigns between 2008 and 2009.
We used the 1.3m telescope at the Skinakas observatory (Crete), the 0.84 m telescope at the OAN (San Pedro
Martir, Mexico) and the 1.5 m telescope at the 
Observatorio de Sierra Nevada (OSN) located in Granada, Spain. 
A total of 945 datapoints in 29 nights and 497 datapoints in 22 nights were collected for the stars
4U\,0115+63 and SAX\,J2103.5+4545 respectively.
The data were obtained primarily in the Johnson V.

\section{Results}

The target 4U\,0115+63 showed variations in 2008, with a frequency of 3.3~\cd\ and an amplitude of 18 mmag.
In 2009, 2 significant frequencies at 3.33 and 1.60~\cd, with amplitudes of 14 mmag and 7 mmag respectively
were detected in the Skinakas data.
As seen in the Be stars observed with the \corot\ mission (eg. \cite{Gutierrez}), the relationship between the
two largest frequencies is very close to 2:1.
The main frequency 3.33~\cd\ is too high to be the rotational frequency.
If we assume that the frequency 1.60~\cd is the rotational frequency, the Be star should rotate at critical
velocity. Therefore, these variations are most probable due to NRP. 

A frequency at 2.23~\cd\ with an amplitude of 4 mmag was found in the light curve of SAX\,J2103.5+4545.
Using the radius and mass of a star with spectral type B0 derived by \cite{Vacca}, and assuming that the star
is rotating at break-up velocity, we calculate a rotational frequency of 1.6~\cd.
Therefore, the frequency 2.23~\cd\ is attributed to non-radial pulsations.

\begin{figure}[ht]
\begin{center}
 \includegraphics[width=6.6cm]{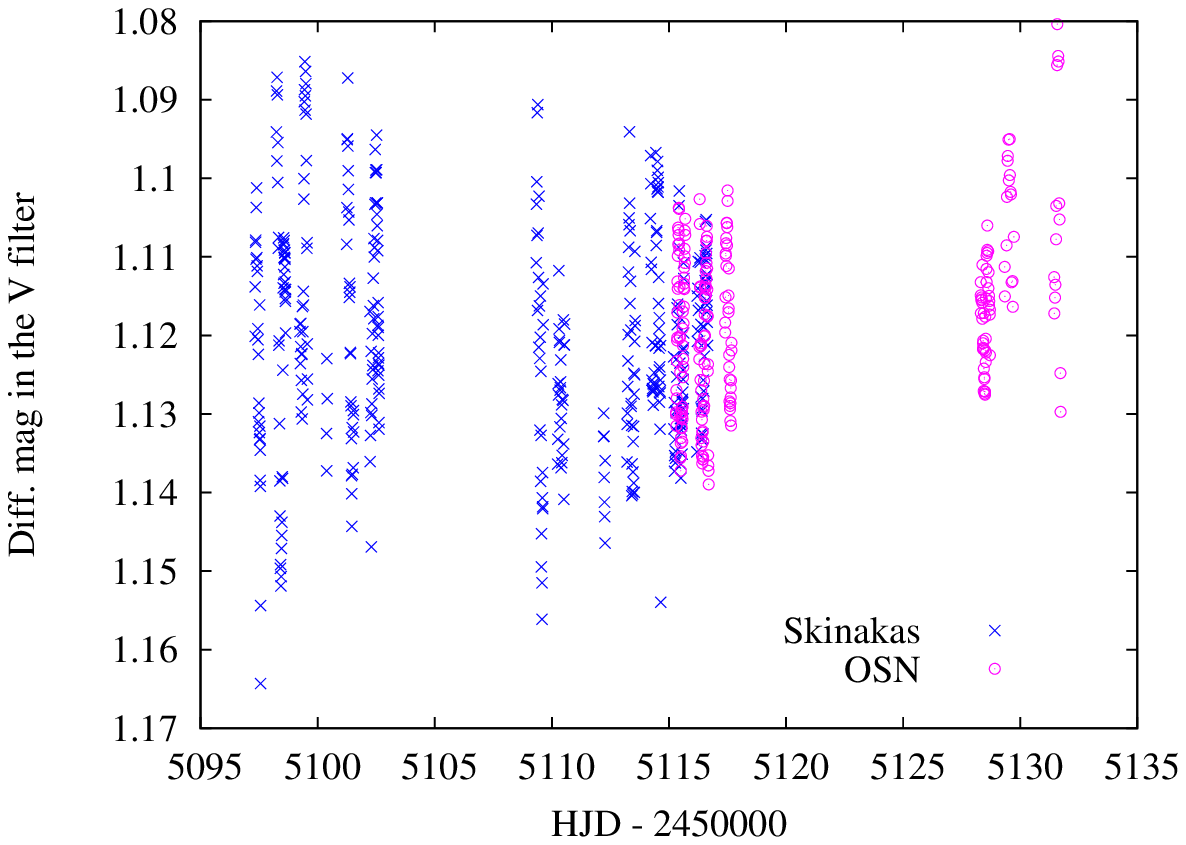} 
\includegraphics[width=6.6cm, bb=54 360 558 720]{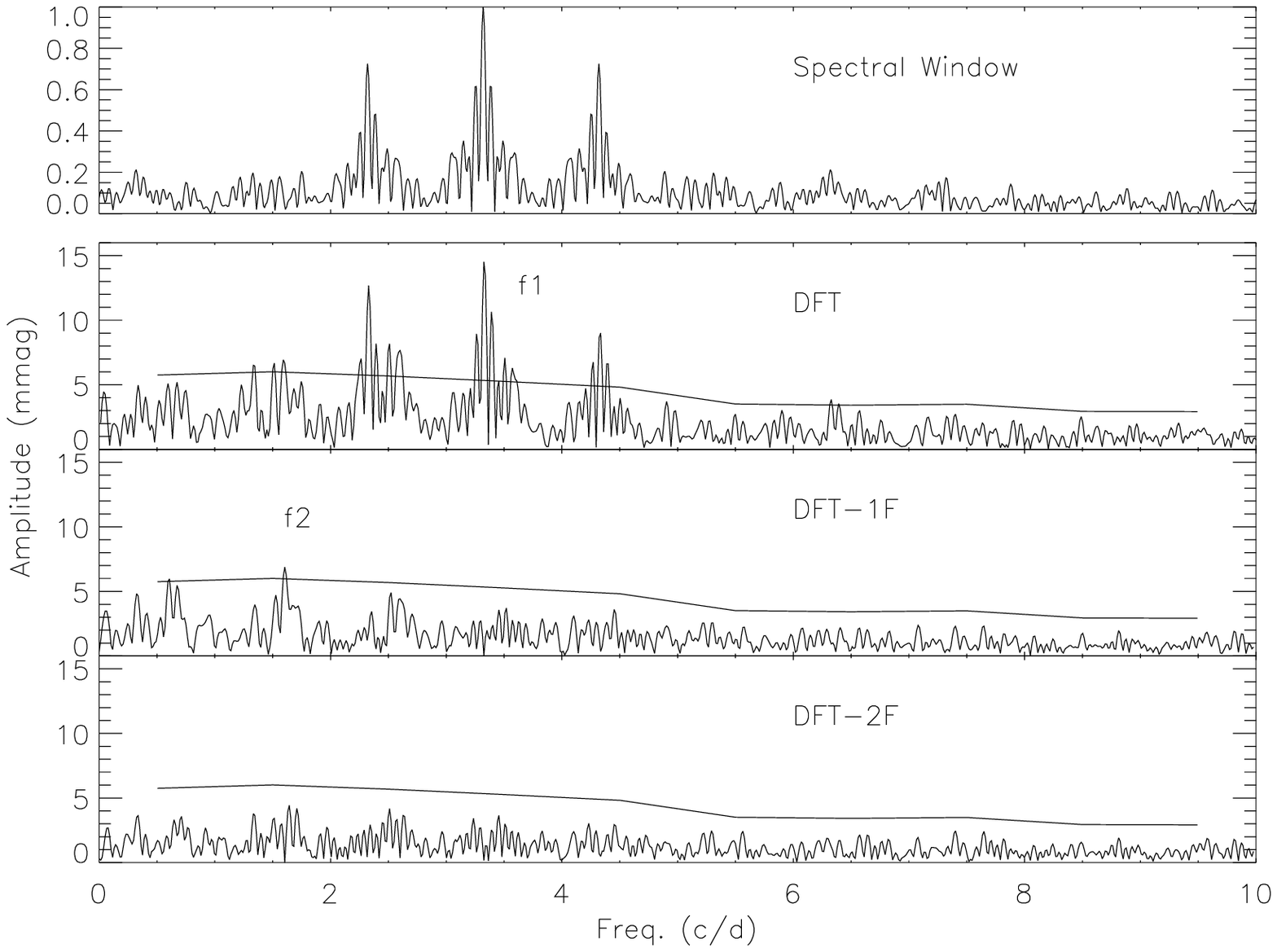} 
 \caption{\textbf{Left}: The Skinakas (blue crosses) and the OSN data (purple circles) for the star
4U\,0115+63.
\textbf{Right}: The subsequent periodograms of the Skinakas data for the star 4U\,0115+63. The horizontal line
represents the 3.5 signal-to-noise level}
   \label{fig2}
\end{center}
\end{figure}


\end{document}